\documentclass[aps,prl,10pt, twocolumn,showpacs,groupedaddress,floats]{revtex4-1} 
\usepackage{amssymb,graphicx,color,amsmath,xfrac,bm,url,oldstyle,soul}
\usepackage[mathletters]{ucs}
\usepackage[utf8x]{inputenc}
\usepackage[english]{babel}

\definecolor{lightblue}{rgb}{0.17,0.39,1}
\definecolor{lightgreen}{rgb}{0.67,0.81,0.08}
\definecolor{lightred}{rgb}{1,0.05,0.52}

\newcommand{\brac}[1]{\left[#1\right]}              

\newcommand{\av}[1]{\left\langle#1\right\rangle}
\newcommand{\hide}[1]{{}}							
								
\renewcommand{\bold}[1]{{\bf{#1}}}

\renewcommand{\S}{{\bf S}}

\begin{document}

\title{Local magnetic moments due to loop currents in metals} 

\author{Arkady Shekhter}
\email[Email address: ]{arkadyshekhter@gmail.com}
\affiliation{Los Alamos National Laboratory, Los Alamos, New Mexico 87545, USA}
\author{Chandra M. Varma}
\email[Email address: ]{chandrav@ucr.edu}
\affiliation{Physics Department, University of California, Berkeley, California 94704, USA}
\altaffiliation{Recalled Professor} 
\affiliation{Physics Department, University of California, Riverside, California  92521, USA}
\altaffiliation{Emeritus Distinguished Professor} 

\begin{abstract}
We present Hartree-Fock calculations on a simple model to obtain the conditions of formation of local magnetic moments due to loop-currents $L_o$ and spin-loop currents $L_s$  and compare them to the conditions of formation of local spin-moments $M$ which were given long ago in a similar approximation by Anderson. A model with three degenerate orbitals sitting on an equilateral triangle, with on-site  and nearest-neighbor repulsions $U$ and $V$ respectively, and inter-site kinetic energy, hybridizing with conduction electrons with a parameter $Δ$ is investigated. $L_o$ and $L_s$ are promoted by large $V/Δ$ and their magnitude is relatively unaffected by $U/Δ$. Spin-magnetic moments $M$ promoted by large $U/Δ$ on the other hand are adversely affected by $V/Δ$. In this model, $L_o$ for $V$ multiplied by the number of neighbors is approximately the same as the $M$ promoted by $U$ in Anderson's local model for $M$. $L_o$ and $L_s$ are degenerate if exchange interactions and Hund's rule are neglected but $L_o$ is favored when they are included. Many of the qualitative results are visible in an expression for the Hartree-Fock ground state energy derived as a function of small $L_o, L_s$ and $M$. Numerical minimization of the Hartree-Fock energy is presented for larger values. We also briefly discuss the connection and differences of the interaction generated orbital currents and spin-currents discussed here and generalized to a lattice with the topological states in metals and semi-conductors.
\end{abstract}
\maketitle
\date{Today}

\section{Introduction}
A number of metallic compounds and some  magnetic insulators have presented evidence recently of broken time-reversal and associated changes in lattice symmetry associated with loop-currents. These include the cuprate high temperature superconductors \cite{Bourges2021, Kaminski-diARPES, Hsieh-SHG-Srcuocl, Hsieh2017} in their so-called pseudogap phase  and some metallic kagome compounds \cite{Kagome_STM_2021, Kagome_Kerr2022}. The insulating state in cuprates has also shown evidence for such states \cite{Hsieh-SHG-Srcuocl} as well as the closely related insulating and conducting states in Sr$_2$(Ir-Rh)O$_4$ \cite{HsiehSrRh, BourgesSr2Ir}. For the cuprates, such states were predicted \cite{cmv1997, simon-cmv} to occur at the boundary to the marginal Fermi-liquid \cite{CMV-MFL, Aji-V-qcf1} phase  with its $T$-linear resistivity \cite{Varma-rmp2020} and singular Fermi-liquid \cite{Varma_2016} anomalies in other properties. Such states have also been predicted for bi-layer graphene \cite{Zhu-A-V2013} and a closely related state has been suggested for twisted bi-layer graphene \cite{Zaletel2020, Berg2021}. The orbital moments due to loop-currents are hard to detect experimentally.  It is quite likely that many other metals and correlated insulators also have such moments. For example, the recently discovered \cite{Kalisky2022} spontaneous vortex state in the superconducting state in layers of TaS$_2$ with field applied and then removed below a specific temperature in the normal state may be due to such a normal state. If described by a  time-reversal breaking vector order parameter such states are expected to have mobile vortices in the vector field generated by an applied magnetic field \cite{CMV2022_R_H}.

Much insight was gained in understanding metallic spin-magnetism through first understanding the formation of local magnetic moments \cite{Friedel_Magmom, PWA1961} on d-orbitals with a large local on-site interaction compared to the hybridization energy with non-interacting s-electrons. One can then study long-range order and other phenomena by considering interactions between the local moments.  This is a strong coupling approach, as distinct from the Stoner-type weak-coupling theory. In this paper, we investigate the formation of magnetic moments through loop-current order in the simplest system that can support such states: three degenerate states on a triangle of sites with on-site and nearest neighbor interactions hybridizing with a continuum of non-interacting states. We do this by following the Hartree-Fock method used by Anderson in the local spin-moment problem \cite{PWA1961}. Although our calculation is instructive, there are several other more subtle features in the correlations in the loop-current states in a lattice. We refer the reader to the results of a detailed variational Monte Carlo calculation \cite{Weber-Giam-V} for them.

It should be appreciated that the magnitude of the orbital moment can be considerable. Suppose there is an electron going around in a circle of radius $a$ and speed $v$. Its classical magnetic moment is $M = π a^2~\mathcal{I} $, where $\mathcal{I}$ is the current going around the wire. If we take that there are $n$ electrons in the wire (i.e., in its volume $(2π a)\,(π r_0^2)$, where $r_0$ is the radius of the cross-section of the wire), $M = (1/2)n(v/c)\,ea$ which for $n =1,a = 2 \text{Å}$ and $v  = 10^8$ cm/sec, the velocity in graphene, is about $1.5 × 10^{-20}$ ergs/Gauss, i.e., close to $1.5 ~ μ_{\text{B}}$. The velocity will often be lower and the density of electrons in the loop-current order may be much less than 1 per unit-cell. In polarized neutron scattering experiments in cuprates \cite{Bourges2021}, a magnitude of (staggered moment) of about $0.1 μ_{\text{B}}$ per unit-cell is deduced. The orbital moments are generated by condensing part of the spatially dependent orbital magnetic susceptibility. The integral of the spectral weight over frequency and momenta of the orbital magnetic fluctuations in the incoherent weight in a noninteracting model of metals is comparable to a similar quantity for spin fluctuations.

\section{The Model}
We consider the model   given by the Hamiltonian 
\begin{align}
H=H_s + H_d +H_{s-d} \,.
\end{align}
$H_s$ is the Hamiltonian for free electrons
\begin{align}
H_s = ∑_{{\bf k} σ} (ϵ_{{\bf k} σ} - μ) c^+_{{\bf k} σ}c_{{\bf k} σ} \,.
\end{align}
$H_d$ describes three orbitals with Kramers' degeneracy on three atoms sitting on an equilateral triangle immersed in a metal of non-interacting electrons. It consists of the noninteracting part $H_d'$ and the part $H_d^{''}$ with onsite and nearest-neighbor interactions,
\begin{align}
H_d' =& ϵ_d ∑_{i = 1,2,3, σ} d_{i σ}^+d_{i σ} + t_{dd} ∑_i d^+_{i σ}d^+_{i+1 σ} + h.c., \\
\label{local-interaction}
H_d^{''} =& U ∑_i n_{i ↑} n_{i ↓} + V ∑_i n_i n_{i+1} \,,
\end{align}
where $n_i = n_{i ↑} +  n_{i ↓}$  and $n_{i σ} = d_{i σ}^+d_{i σ}$.
$H_{sd}$ is the hybridization
\begin{align}
H_{sd} = ∑_{i \bf k σ} t_{i {\bf k}} d_{iσ}^+ c_{\bf k σ} + h.c.
\end{align}
The problem is best discussed  in  the basis of the eigenstates of angular momenta defined with respect to the center of the triangle
\begin{align}
D_{ℓ σ} = \frac{1}{\sqrt{3}} ∑_{n} d_{n σ} e^{i ω_ℓ n}.
\end{align}
where $ℓ = 0, ± 1$ and $ ω_{ℓ} = \frac{2π}{3} ℓ$. When an electron is in the state $ℓ = ± 1$, the triangle carries a clockwise or anticlockwise loop current. In the new basis
\begin{align}
H_{sd} = ∑_{ℓ} t_{ℓ} D^+_{ℓ σ} c_{ℓ k, σ} + h.c.
\end{align}
conserving both angular and spin momenta. $c_{ℓ k, σ}$ are operators for hole states with angular momenta $ℓ = 0, ± 1$ with respect to the center of the triangle. 
The Hamiltonian $H_d'$ is diagonal in the angular momentum basis
\begin{align}
H_d' = ∑_{ℓ σ} ϵ_{d ℓ} n_{ℓ σ}. 
\end{align}
The energy levels, including the shift due to the hybridization with the Fermi-sea,
 are  $ϵ_{d ℓ} = ϵ_d + 2 t_d \cos ω_{ℓ}$. Note that the relative positions of the singlet $ℓ =0$ level and the doublet $ℓ = ± 1$ are given by the sign of $t_d$. We will assume that the doublet is below the singlet, as is the case for $t_d > 0$.

Finally we write the interactions in the new basis. To demonstrate the basic idea, we first consider spinless electrons.
In this case, the interactions in the angular momentum basis are
\begin{align}
H_d{''} = V  n_0 (n_{+1} + n_{-1}) + V n_{+1}n_{-1}\,.
\end{align}
The first term serves to shift the energy difference between the orbital singlet and the orbital doublet. The last term introduces the repulsion between $ℓ = ± 1$ states in the Hartree approximation. If this were the only interaction term in the model, the problem would have a one-to-one correspondence with the Hartree-Fock repulsion between up- and down-spins due to the on-site interactions in Anderson's calculation \cite{PWA1961}.  

We now go back to the spinful model, which is a bit more involved. We must consider the Hartree-Fock energy $E_{ℓ, σ}$ of the states ($ℓ =0, σ = ± 1); (ℓ = ± 1, σ = ± 1).$ Terms in $E_{ℓ, σ}$ that are linear in $n_{ℓ σ}$ give the bare energies, and those due to interactions are quadratic in $n_{ℓ, σ}$. $n_{ℓ σ}$ are determined variationally from the Hartree-Fock ground state energy $E_0({n_{ℓ σ}})$ which can be written as the sum of a potential energy $P({n_{ℓ σ}})$ and a kinetic energy $K({n_{ℓ σ}})$:
\begin{align}
\label{HFenergy}
E_0(n_{ℓ σ}) = P(n_{ℓ σ}) + K(n_{ℓ σ})\,.
\end{align}
For a constant density of states of conduction electrons $ν$ per spin, the kinetic energy is, just as in Anderson's calculations 
\begin{align}
K({n_{ℓ σ}}) = -  ∑_{ℓ σ} \frac{Δ_{ℓ}}{π} \ln \sin (π n_{ℓ σ}) \, .  
\end{align}
$Δ_{ℓ} = π |t_{ℓ}|^2 ν$ is the width of the levels due to hybridization with the itinerant electrons.

The (coupled) variational equations are 
\begin{align}
Δ_{ℓ} \cot (π n_{ℓ σ}) = \frac{∂ P(n_{ℓ σ})}{∂ n_{ℓ σ}}.
\end{align}
The potential energy may be written as the part for $n_{0 σ}$ and the part for $n_{1 σ}$:
\begin{align}
P(&n_{0 σ}) = ∑_{σ} ϵ_{d0} n_{0 σ} \notag\\ 
  +& \frac{U+2V}{3} n_{0 σ} N_{-σ}  + V n_{0 σ} N_{σ} +  \frac{U+2V}{6} n_{0 σ}n_{0 -σ}, \\
  P(&n_{± 1 σ}) =∑_{σ} ϵ_{d1} N_{σ}   \notag\\
   &\qquad + \frac{U+2V}{6} N_{σ} N_{-σ}  + V n_{1 σ} n_{-1 σ} \,, 
     \label{potential-orbital-basis} 
   \end{align}
   where $N_{σ} ≡ (n_{1 σ} + n_{-1 σ})$.
   
  We introduce ground state expectation values
   \begin{align}
   n ≡ ∑_{σ} \av{n_{0 σ}}\,, \qquad m ≡ ∑_{σ} \av{σ n_{0 σ}}\,,
   \end{align} 
as the  occupation and total spin-moment of the state with $ℓ =0$. We also define
   \begin{align}
   N ≡& ∑_{ℓ σ} \av{n_{ℓ σ}}\,, \qquad\;\; M ≡ ∑_{ℓ σ} σ \av{n_{ℓ σ}}, \notag\\
   L_o ≡& ∑_{ℓ σ} ℓ  \av{n_{ℓ σ}}\,, \qquad L_s  ≡ ∑_{ℓ σ} ℓ σ \av{ n_{ℓ σ}} \,, 
   \end{align}
   as, respectively, the total occupation, the total spin, the total orbital moment and the total spin-current.
   
   To simplify the analysis, let us consider the case where the $ℓ =0$ level is well above the chemical potential and therefore    $n_{0 σ} =0$. (Similar results pertain when it is well below the chemical potential and is filled with both spins.) The potential energy then is
   \begin{align}
   \label{pot}
   &P(N,M, L_o,L_s) = ϵ_{d1} N \nonumber \\
   +& \frac18\brac{\frac{2U + 7V}{3} N^2 - \frac{2U + V}{3} M^2 - V(L_o^2 + L_s^2) }\,.
   \end{align}
  To get an idea of the different phases as a function of the parameters, let us fix the total occupation $N$ and expand the kinetic energy
  as a function of small $M, L_o$ and $L_s$. This gives
  \begin{align}
  \label{kin}
  &\frac{K}{Δ} = c_0 + c_2(M^2 +L_o^2 +L_s^2) + c_3 M L_o L_s \notag\\ 
      +& c_4 (M^4+L_o^4 +L_s^4 + 6 L_o^2M^2 +6 L_s^2 M^2 + 6 L_o^2L_s^2),
    \end{align}
    where
    \begin{align}
    c_0 =& - \frac{1}{π} \ln \sin \frac{π N}{4}, 
    \qquad 
    c_2 = \frac{π}{8} \csc \frac{π N}{4} \,, \notag\\  
    c_3 =& - \frac{π^2}{8}  \cot  \frac{π N}{4}  \csc^2  \frac{π N}{4} \,, \notag\\ 
     c_4 =& \frac{1}{24} \; \frac{π^3}{16} \;\; \frac{\frac{1}{2} + \cos^2 \frac{π N}{4} }{\sin^4 \frac{π N}{4} } \,.
    \end{align}
    Several noteworthy features are already apparent. Obviously  the potential energy favors non-zero values of $M,L_o$ and $L_s$ while the kinetic energy favors $0$. In the potential energy as well as the kinetic energy, one can interchange $L_o$ and $L_s$. Potential energy allows both $\sqrt{L_o^2+L_s^2}$  and $M$ to be non-zero. Since $c_4 > 0$, the kinetic energy favors either $M$ or $L_o$ or $L_s$. The degeneracy between $L_o$ and $L_s$ together with competition  with non-zero $M$ from the cubic term proportional to $c_3$, which has a positive sign above for $2<N<4$ introduces interesting features in numerical calculations. The degeneracy is removed as discussed below when we introduce exchange interactions.
    
In the vicinity of empty, $N ≈ 0$, and full, $N ≈ 4$, $c_2$ is very large  and therefore $M, L_o$, and $L_s$ are all $0$. Near half filling, $N ≈ 2$, $c_2 $ is near its minimum value $1/2$. So this is the most favorable value for all the polarizations. Near this value of $N$, $c_4 ≈ 1/10$ and $c_3 ≈ 0$. Comparing coefficients in Eqs. (\ref{kin}) and (\ref{pot}), one easily finds that for $V > c_2 Δ$, and $(2U+V)/3 < c_2 Δ$, the bi-linear terms favor non-zero loop-currents $L_o, L_s$ while $M =0$.  At precisely $N =2$, where $c_3 =0$, we get $L_o^2 = \frac{2}{5}(V/Δ - c_2)/c_4$. Away from half filling, cubic terms  allow ``first-order'' changes in the values of $M,L_o$, and $L_s$. Magnetic moments enter linearly in the cubic terms and lead to $ M ∝ L^2$. 

For small values of $c_4$, values of $M,L_o$, and $L_s$ quickly reach $1$ with increasing  $U/Δ, V/Δ$ and the expansion above becomes invalid. We therefore have performed a numerical minimization of the Hartree-Fock energy for fixed $N$ and varying $M,L_o$, and $L_s$. The results are shown in Figs.~\ref{Fig:HF1}~and~\ref{Fig:HF2}.

\subsection{Orbital moment due to loop-current $L_o$ versus spin-loop current $L_s$}

In the Hartree-Fock effective Hamiltonian of the model with energy given by Eq.~(\ref{HFenergy}) and subsequent equations following from it, there is an ``accidental'' symmetry, which is also manifest in the low order parameter expansion for the energy, Eqs.~(\ref{pot}) and (\ref{kin}). The accidental character of this symmetry is obvious because the state with finite $L_o$ has the same energy as the state with finite $L_s$ despite the fact that these states have different symmetries --- $L_o$ is odd in time reversal while $L_s$ is even. The symmetry of the energy under the exchange of orbital and spin currents,  $L_o$ and $L_s$  is introduced due to the de-couplings of the biquadratic operators of spin and angular momentum in the Hartree-Fock approximation which in an exact theory would not be allowed. 

This matter is resolved by including exchange interactions. In the limit of small inter-site repulsion and large on-site repulsion, $V < t ≪ U$ we can include the exchange Hamiltonian, 
\begin{align}
\label{kekule}
H_{\text{exch}} = J({\bf S}_1⋅ {\bf S}_2 + {\bf S}_2 ⋅ {\bf S}_3 + {\bf S}_3⋅ {\bf S}_1)
\end{align}
where the subscripts are site indices. The value of the exchange $J ≈ {2t^2}/({U-V})$. 
Neglecting the $ℓ =0$ state, the exchange Hamiltonian in the orbital states basis is
\begin{align}\label{eq:exch}
H_{\text{exch}} =  - J  \S_{ℓ =-1}⋅\S_{ℓ =1}  
\end{align} 
This includes the fact that when a pair of electrons are in the same orbital, the Pauli principle ensures a spin-singlet state and when they are in different orbitals, Hund's rule favors a triplet state. We lose the $SU(2)$ symmetry by the Hartree-Fock approximation on (\ref{eq:exch}) and write it in terms of only z-components of spin as done above for the rest of the interactions. This exchange energy  shifts the expectation values of the energy of the states described in terms of occupation numbers Eq.~(\ref{HFenergy}), $H_{\text{exch-HF}} =  - \tilde{J} ( M^2 - L_s^2 ) $, where $\tilde{J} = J/4$ and $L_s =  \av{S_{ℓ =1}} - \av{S_{ℓ =-1}} $, $M =  \av{S_{ℓ =1}} + \av{S_{ℓ =-1}} $. Equation~(\ref{kekule}) applies only for $V < t ≪ U$ where the formation of orbital current is not favorable. In the opposite limit of large inter-site repulsion $V$, the loop-orbital states $ℓ =±1,0$, rather than the localized site-states on the triangle (Eq.~\ref{kekule}), are a good starting point, as discussed above. The local interaction Eq.~(\ref{local-interaction}) contains non-Hartree-Fock terms $δ H_2 = (V-U)/6 \, [ S_{x,-1} S_{x,+1} +  S_{y,-1} S_{y,+1} ]$ omitted in the Hartree-Fock energy Eq.~(\ref{potential-orbital-basis}). These can be rewritten as $δ H_2 = (V-U)/6 \, [ \S_{-1} \S_{+1} -  S_{z,-1} S_{z,+1} ]$, which can be estimated as $δ E_2 =J \, [ (1/4)M^2 + (1/4) J_s^2 ]$.  The first term can be absorbed in the existing expression for the energy, Eq.~(\ref{pot}). The second term, 
\begin{align}\label{eq:exchHF}
H_{\text{exch-HF}} =   \tilde{J} L_s^2   \,, 
\end{align}
lifts the accidental degeneracy in the Hartree-Fock approximation between states with nonzero orbital loop current or spin current states and it favors the orbital magnetic moment over the spin-current moment. In the limit of large $V$, parameter $J$ has to be found from a complete SU(2) analysis.

\section{Numerical results}

\begin{figure*}[ht!!!!!!!!!!]
 	\includegraphics[width= 0.7\textwidth]{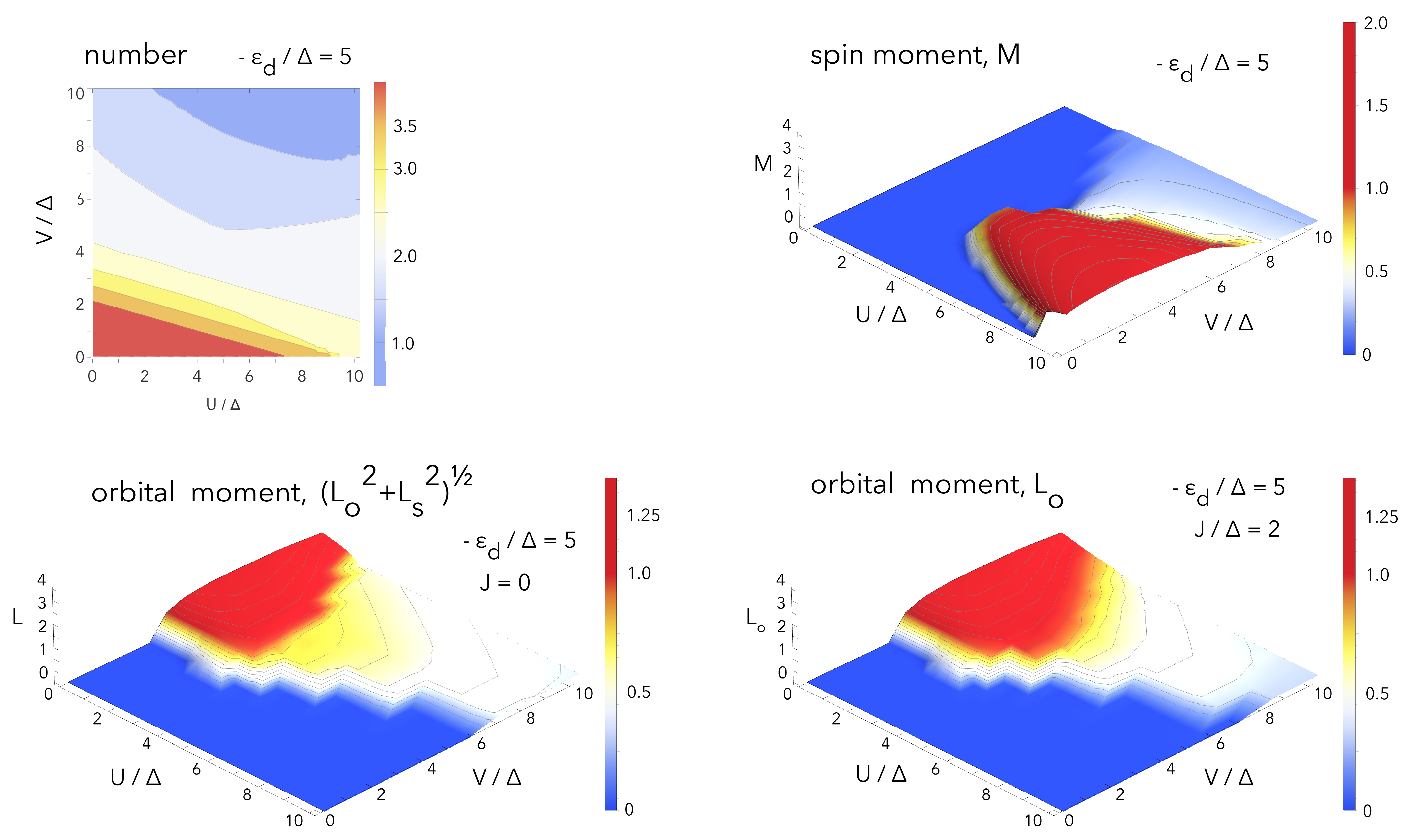}
\caption{Top left: Total charge $N$ as a function of interaction $U/Δ$ and $V/Δ$ for the orbitally degenerate non-interacting levels with $ϵ/Δ$ such that $N$ would be $4$ for non-interacting electrons. Top right:  Spin magnetic moment as a function of the interaction parameters $U/Δ$ and $V/Δ$ for the same $ϵ/Δ$ as in the top left. The bottom left and right present $\sqrt{L^2_o + L^2_s}$ for the case where the exchange interaction $J =0$. As explained in the text, numerical calculations give either $L_s \ne 0$ or $L_o \ne 0$ due to numerical noise but without any ambiguity for $\sqrt{L^2_o + L^2_s}$. Including $J \ne 0$ removes the degeneracy and favors $L_o$. That value is shown in the bottom right.}
	\label{Fig:HF1} 
\end{figure*}

The numerical calculations are also done with the $ℓ =0$ level far above the chemical potential so that it is un-occupied. In  Fig.~(\ref{Fig:HF1}), $ϵ_d/Δ = -5$ is chosen so that the non-interacting triangle would have both $ℓ = ± 1$ fully occupied and the ground state has four electrons. The panel on the left shows the ground state charge $N$ as a function of $U/Δ$ and $V/Δ$. As expected, the number goes down with these repulsions. The interesting thing is the stability of $N=2$ over a large range of the interaction parameters. This can be traced to the reduction in energy due to the maximum possible development of the spin moment or orbital moment for $N=2$. The spin moment $M$, the orbital moment $L_o$, and the spin current are shown in plots with varying $U/Δ$ and $V/Δ$. The spin-moment, shown by the figure in the top right using a small $V$, follows Anderson's result \cite{PWA1961}, except that the value of $M$ is twice larger because of the orbital degeneracy. The decrease in the local moment at large $V$ and the ``almost-steps'' in the topography should be noted; this may be attributed to lifting of the orbital degeneracy by $V$ and the consequent decrease in $N$ noted in the top left-hand panel. But for small $U$, a weak growth of the spin moment with $V$  is also noteworthy.  

\begin{figure*}[ht!]
	 \includegraphics[width=0.7\textwidth]{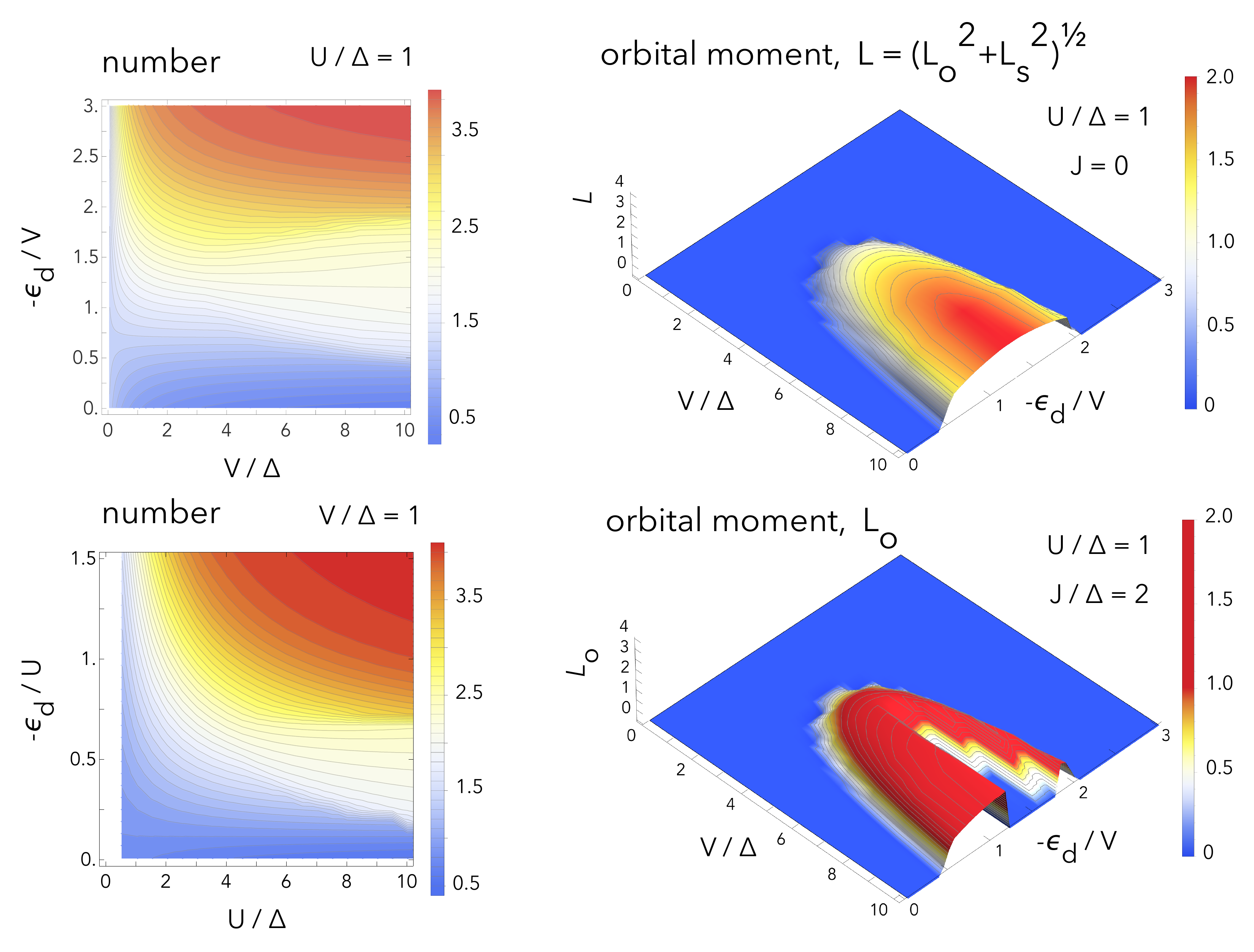} 
 \caption{Left: The top figure shows the total  charge for a modest value of $U/Δ$ in the $ϵ_d/V$ and $V/Δ$ plane, while the bottom figure shows the same for a fixed modest $V/Δ$ in the $ϵ_d/U$ and $U/Δ$ plane. Right: The top figure shows the result for $J=0$ when $L_s$ and $L_o$ are degenerate in our approximation. The bottom shows $L_o$ for finite $J$ for which $L_s =0$.}  
 \label{Fig:HF2}
 \end{figure*}

We first show the calculations for $L_o$ and $L_s$ without including the exchange interaction parameterized by ${\tilde J}$. As discussed in the previous section, the Hartree-Fock approximation for ${\tilde J} = 0$ introduces an ``accidental'' degeneracy between $L_o$ and $L_s$. This is seen in the numerical results which give the state realized to be either $L_o \ne 0$ with $L_s =0$ or vice-versa depending on the numerical noise. One or the other is favored due to the repulsions seen already in the low moment expansions of the energy above in Eq.~(\ref{kin}) due to the cubic term. On the other hand we find no ambiguity when we plot $\sqrt{L_o^2 +L_s^2}$. This is what is shown in the bottom left in Fig.~(\ref{Fig:HF1})  which is for  ${\tilde J}  =0$.  On the bottom right, we plot the results for $\tilde{J}$ with its value fixed by the other parameters as in Eq.~(\ref{eq:exch}). $L_o$ is favored, although numerical noise appears to give also a much smaller value of $L_s$ for some range of parameters.

 The figures at the bottom right in Fig.~(\ref{Fig:HF1})  gives the orbital magnetic moment with a maximum amplitude $1$ at large $V$ corresponding to the complete lifting of the degeneracy of the $ℓ = ± 1$ state. The value of $V$ required for this is about 1/2 the value of $U$ for a maximum spin moment at small $V$, due to the fact that the effective value of $V$ to generate the orbital moment gets multiplied by the number of neighbors, two in this case. One finds that the modest value of  $U$ does not decrease the orbital moment. This is due to the fact that both a spin-loop-current and a charge-loop-current are favored for large $V$. At large values of $U$, the change of $N$ is responsible for the decrease of $L_o$ in a step-like fashion. This is due to the fact that a charge-loop-current is favored for large $V$. We have already discussed following Eq.~(\ref{kin}) the competition of $L$ with $M$ due to the fourth order repulsive terms coupling $L_o$ and $M$ proportional to $c_4$. Further details of the numerical results obtained  are illustrated in Fig.~\ref{Fig:HF2}.

\section{General remarks}

For simplicity we considered atoms on an equilateral triangle so that the unperturbed states could be labeled by their angular momentum. Then one need only consider interactions which lift the Kramers' degeneracy between the $± 1$ angular momentum so that if the chemical potential is not above the newly diagonalized states, a moment due to loop-current is realized. This also stresses the connection of the sign of the product of the transfer integral between the three atoms and the average density of particles to have the orbital moment.

These considerations can be generalized to ``molecules'' with different symmetries. For problems in the lattice, the patterns which are realized must satisfy the condition that the currents are conserved at each site. In a mean-field approximation, the local order parameter $\av{C}$ added to a real kinetic energy operator corresponds to a complex hopping between $i$ and $j$ for $\av{C}$ purely real and therefore a current for $σ = σ'$ and a spin-current for $σ \ne σ'$. For $\av{C}$ purely imaginary, it represents an excitonic charge transfer which would in general couple also to the lattice and lead to a structural transformation. The identity can be transformed to momentum space and one could do mean-field theory in the weak-coupling approximation to look for all such instabilities. This was the procedure used in Ref. \cite{simon-cmv} for cuprates.

The Hartree-Fock calculations, although revealing, are hardly the complete story for the conditions of occurrence of orbital moments in a lattice. These are considered in more detail in Refs.~\cite{Weber-Giam-V, Weber-Mila}. A primary consideration is that for a metallic state to exist requires charge fluctuations and these cost energy due to repulsive interactions which are usually ameliorated by screening, generally by relatively un-correlated $s-p$ electrons which in turn cost kinetic energy for them. The s-p electrons are often not in the models and sometimes, as in the cuprates and other transition metals to the right of the periodic table, too far in the majority of their density of states above the chemical potential to be effective. A loop-current ordered state allows a coherent screening within the correlated states reducing the kinetic energy to screen. A significant next nearest neighbor interaction, necessary for loop-current states, also requires the absence of significant effects of $s-p$ electrons in actual materials.

There are two other possible states in the triangle when the total charge is $2$: One is an in-homogeneous threefold degenerate state in which two of the sites are occupied and the third is unoccupied. This should probably be called a ``Kekule'' structure. With interactions eliminated to $O(t/(U,V)$ this state is further stabilized by the exchange energy so that a singlet bond is formed on a pair of sites with the third site unoccupied. Such a state would also persist for $N$ close to $2$. Consideration of such inhomogeneous states requires considering the $ℓ =0$ state which we have neglected. The spin-current state discussed is also likely to be modified if exchange interactions are considered. The state is one in which the Kekule structure is a linear combination of spin-singlet bonds in the three pairs of sites which should be appropriately called a ``resonating valence bond'' state on a triangle is related to the orbital current state investigated here but for one important difference. Since $SU(2)$ spin-symmetry must be maintained for such a state, a simple Hartree-Fock approximation which is the basis of the results in this paper is not appropriate. Retaining $SU(2)$ symmetry is also essential for a proper theory of the spin-current state in a model including exchange interactions.

Finally, we discuss the connection of the loop-currents generated by interactions in the lattice version \cite{Varma-rmp2020} of the physical states such as discussed in this paper, with the topological states with anomalous Hall effects \cite{Haldane1988}, or anomalous spin-Hall effects \cite{KaneMele2005, NiuRMP2010}. Both kinds of effects depend on having a lattice with a basis (whether in the lattice already or after the symmetry breaking induced by the interactions.) The latter depend crucially on the topological effects of the periodicity of the Brillouin zone in more than one dimension, and the former do not. The Haldane state \cite{Haldane1988} on a hexagonal lattice can be realized only through loop currents due to next nearest neighbor interactions, as explained in Ref. \cite{Zhu-A-V2013} but it has the additional feature that inversion is preserved. Alternate loop-currents on the hexagonal lattice investigated in the same paper are also promoted by interactions but they break inversion. Then the anomalous Hall effect cannot occur. The anomalous spin-Hall currents require spin-orbit coupling which at least in non-relativistic physics should be thought of as an intra-atomic interaction effect and some specific conditions on symmetry at special points in the zone of the conduction and valence band states. However, the spin-currents discussed here together with spin-orbit couplings lead to the same physics if those conditions are satisfied \cite{NiuRMP2010}. In general, the role of electron-electron interactions in stabilizing orbital magnetism is less obvious in the topological arguments in momentum space than in the real space analysis which has been the emphasis in this paper.

{\it Acknowledgements:} 
 Work at Los Alamos is supported by the NSF through DMR-1644779 and the U.S. Department of Energy. A.S. acknowledges support from the DOE/BES ``Science of 100 T'' grant. This work was performed in part at Aspen Center for Physics, which is supported by National Science Foundation grant No. PHY-1607611.


\begin{thebibliography}{22}

\bibitem{Bourges2021} P. Bourges, D. Bounoua, and Y. Sidis, \emph{Comptes Rendus Physique} \bold{22}, 7 (2021).
\bibitem{Kaminski-diARPES} A. Kaminski, S. Rosenkranz, H. M. Fretwell, J. C. Campuzano, Z. Li, H. Raffy, W. G. Cullen, H. You, C. G. Olson, C. M. Varma, and H. Hochst, \emph{Nature} \bold{416}, 610 (2002). 
\bibitem{Hsieh-SHG-Srcuocl} A. de la Torre, K. L. Seyler, L. Zhao, S. Di Matteo, M. S. Scheurer, Y. Li, B. Yu, M. Greven, S. Sachdev, M. R. Norman, and D. Hsieh, \emph{Nature Physics} \bold{17}, 777 (2021).
\bibitem{Hsieh2017} L. Zhao, C. A. Belvin, R. Liang, D. A. Bonn, W. N. Hardy, N. P. Armitage, and D. Hsieh, \emph{Nature Physics} \bold{13}, 250  (2016).
\bibitem{Kagome_STM_2021} K. Jiang, T. Wu, J.-X. Yin, Z. Wang, M. Z. Hasan, S. D. Wilson, X. Chen, and J. Hu, National Science Review, nwac199 (2021).
\bibitem{Kagome_Kerr2022} Y. Xu,  Z. Ni, Y. Liu,  B. R. Ortiz, S. D. Wilson, B. Yan,  L. Balents, L. Wu, arXiv:2204.10116, (2022).
\bibitem{HsiehSrRh} L. Zhao, D. H. Torchinsky, H. Chu, V. Ivanov, R. Lifshitz, R. Flint, T. Qi, G. Cao, and D. Hsieh, \emph{Nature Physics} \bold{12}, 32 (2016).
\bibitem{BourgesSr2Ir} J. Jeong, Y. Sidis, A. Louat, V. Brouet, and P. Bourges, \emph{Nature Communications} \bold{8}, 15119 (2017).
\bibitem{cmv1997} C. M. Varma, \emph{Phys. Rev. B} \bold{55}, 14554 (1997).
\bibitem{simon-cmv} M. E. Simon and C. M. Varma, \emph{Phys. Rev. Lett.} \bold{89}, 247003 (2002).
\bibitem{CMV-MFL} C. M. Varma, P. B. Littlewood, S. Schmitt-Rink, E. Abrahams, and A. E. Ruckenstein, \emph{Phys. Rev. Lett.} \bold{63}, 1996 (1989).
\bibitem{Aji-V-qcf1} V. Aji and C. M. Varma, \emph{Phys. Rev. Lett.} \bold{99}, 067003 (2007).
\bibitem{Varma-rmp2020} C. M. Varma, \emph{Rev. Mod. Phys.} \bold{92}, 031001 (2020). 
\bibitem{Varma_2016} C. M. Varma, \emph{Reports on Progress in Physics} \bold{79}, 082501 (2016).
\bibitem{Zhu-A-V2013} L. Zhu, V. Aji, and C. M. Varma, \emph{Phys. Rev. B} \bold{87}, 035427 (2013).
\bibitem{Zaletel2020} N. Bultinck, E. Khalaf, S. Liu, S. Chatterjee, A. Vishwanath, and M. P. Zaletel, \emph{Phys. Rev. X} \bold{10}, 031034 (2020).
\bibitem{Berg2021} J. S. Hofmann, E. Khalaf, A. Vishwanath, E. Berg, and J. Y. Lee,  arXiv:2105.12112 (2021).
\bibitem{Kalisky2022} E. Persky, A. V. Bjørlig, I. Feldman, A. Almoalem, E. Altman, E. Berg, I. Kimchi, J. Ruhman, A. Kanigel, and B. Kalisky, \emph{Nature (London)} \bold{607}, 692 (2022). 
\bibitem{CMV2022_R_H} C. M. Varma, \emph{Phys. Rev. Lett.} \bold{128}, 206601 (2022).
\bibitem{Friedel_Magmom} J. Friedel, Canadian \emph{Journal of Physics} \bold{34}, 1190 (1956).
\bibitem{PWA1961} P. W. Anderson, \emph{Phys. Rev.} \bold{124}, 41 (1961).
\bibitem{Weber-Giam-V} C. Weber, T. Giamarchi, and C. M. Varma, \emph{Phys. Rev. Lett.} \bold{112}, 117001 (2014).
\bibitem{Weber-Mila} C. Weber, A. Lauchli, F. Mila, and T. Giamarchi, \emph{Phys. Rev. Lett.} \bold{102}, 017005 (2009).
\bibitem{Haldane1988} F.D.H. Haldane, \emph{Phys. Rev. Lett.} \bold{61}, 2015 (1988).
\bibitem{KaneMele2005} C.L. Kane and E.J. Mele, \emph{Phys. Rev. Lett.} \bold{93}, 197402 (2005).
\bibitem{NiuRMP2010} D. Xiao, M.-C. Chang and Q. Niu, \emph{Rev. Mod. Phys.} \bold{82}, 1959 (2010)

\end{thebibliography}
\end{document}